# THE ENERGY DISTRIBUTION OF ATOMS IN THE FIELD OF THERMAL BLACKBODY RADIATION

Fedor V.Prigara

*Institute of Microelectronics and Informatics, Russian Academy of Sciences, 21 Universitetskaya, 150007 Yaroslavl, Russia*

Using the principle of detailed balance and the assumption on the absorption cross-section consistent with available astrophysical data, we obtain the energy distribution of atoms in the field of thermal blackbody radiation and show that this distribution diverges from the Boltzmann law.

The synchrotron interpretation of radio emission from active galactic nuclei encounters some difficulties. In particular, the spherical accretion models with the synchrotron emission are unable to explain the flat or slightly inverted radio spectra of low-luminosity active galactic nuclei [1]. Because of these reasons it makes sense to retry the thermal interpretation of radio emission from active galactic nuclei [2] and to investigate more closely thermal blackbody radiation.

Recently, it was shown that the standard theory of thermal radio emission [3] cannot explain the radio spectra of planetary nebulae at high frequencies without an introduction of indefinite parameters [4]. However, the standard theory does not take into account a stimulated character of thermal radio emission following from the relations between Einstein's coefficients for a spontaneous and induced emission of radiation [5]. A revised theory of thermal radio emission gives the radio spectra both of planetary nebulae and of low-luminosity active galactic nuclei [2,5].

The original Einstein theory of spontaneous and induced emission of radiation did not take into account a line width [6]. Here we show, in particular, that the account for a natural line width essentially changes the conclusions on the wavelength dependence of the absorption and stimulated emission cross-sections as well as the relations between Einstein's coefficients.

Consider two level system (atom or molecule) with energy levels $E_1$ and $E_2 > E_1$ which is in equilibrium with thermal blackbody radiation. We denote as $v_s = A_{21}$ the number of transitions from the upper energy level to the lower one per unit time caused by a spontaneous emission of radiation with the frequency $\omega = (E_2 - E_1)/\hbar$, where $\hbar$ is the Planck constant. The number of transitions from the energy level $E_2$ to the energy level $E_1$ per unit time caused by a stimulated emission of radiation may be written in the form

$$\nu_i = B_{21} B_\nu \Delta \nu = \sigma_{21} \frac{\pi B_\nu}{\hbar \omega} \Delta \nu, \qquad (1)$$

where

$$B_\nu = \hbar \omega^3 / (2\pi^2 c^2)(\exp(\hbar \omega / kT) - 1) \qquad (2)$$

is a blackbody emissivity (Planck's function), $\sigma_{21}$ is the stimulated emission cross-section, $T$ is the temperature, $k$ is the Boltzmann constant, $c$ is the speed of light, and $\Delta \nu$ is the line width.

The number of transitions from the lower level to the upper one per unit time may be written in the form

$$\nu_{12} = B_{12} B_\nu \Delta \nu = \sigma_{12} \frac{\pi B_\nu}{\hbar \omega} \Delta \nu. \qquad (3)$$

where $\sigma_{12}$ is the absorption cross-section.

Here $A_{21}$, $B_{21}$ and $B_{12}$ are the coefficients introduced by Einstein [6], the coefficients $B_{12}$ and $B_{21}$ being modified to account for the line width $\Delta \nu$.

We denote as $N_1$ and $N_2$ the number of atoms occupying the energy levels $E_1$ and $E_2$, respectively. The levels $E_1$ and $E_2$ are suggested for simplicity to be non-degenerated.

In the equilibrium state the full number of transitions from the lower level to the upper one is equal to the number of reverse transitions:

$$N_1 \nu_{12} = N_1 B_{12} B_\nu \Delta \nu = N_2 \nu_{21} = N_2 (A_{21} + B_{21} B_\nu \Delta \nu). \qquad (4)$$

The line width $\Delta \nu$ is suggested to be equal to the natural line width [7]:

$$\Delta \nu = A_{21} + B_{21} B_\nu \Delta \nu. \qquad (5)$$

It is assumed that the lower level has a zero energy width, i.e. this level is a ground state.

It follows from the last equation that

$$\Delta \nu = A_{21} / (1 - B_{21} B_\nu). \qquad (6)$$

Substituting this expression in the equation (4), we obtain

$$N_2 / N_1 = B_{12} B_\nu = \sigma_{12} \frac{\pi B_\nu}{\hbar \omega}. \qquad (7)$$

In the limit of high temperatures $T \to \infty$, corresponding to the range of frequencies $\hbar\omega < kT$, the function $B_\nu(T)$ is given by the Rayleigh-Jeans formula

$$B_\nu = 2kT\nu^2/c^2, \qquad (8)$$

where $\nu$ is the frequency of radiation, $\nu = \omega/2\pi$.

Since $B_\nu \to \infty$ when $T \to \infty$, it follows from the equation (6) that the coefficient $B_{21}$ is depending on the temperature $T$ in such a manner that $B_{21}B_\nu < 1$. It is clear that $B_{21}B_\nu \to 1$ when $T \to \infty$. The ratio of frequencies of transitions caused by spontaneous and induced emission of radiation is given by the expression

$$\nu_s/\nu_i = A_{21}/(B_{21}B_\nu \Delta\nu) = (1 - B_{21}B_\nu)/(B_{21}B_\nu). \qquad (9)$$

This ratio is approaching zero when $T \to \infty$, since $B_{21}B_\nu \to 1$. It means that in the range of frequencies $\hbar\omega < kT$ thermal radiation is produced by the stimulated emission, whereas the contribution of a spontaneous emission may be neglected.

It follows from equations (7) and (8) that in the Rayleigh-Jeans range of frequencies the ratio $N_2/N_1$ is given by the formula

$$N_2/N_1 = \sigma_{12}\omega kT/(2\pi\hbar c^2). \qquad (10)$$

An analysis of observational data on thermal radio emission from various astrophysical objects suggests that the absorption cross-section does not depend on a wavelength of radiation and has an order of magnitude of an atomic cross-section, $\sigma_{12} \approx 10^{-15} cm^2$ [2,5]. Only this assumption is consistent with the observed spectra of thermal radio emission from major planets, the spectral indices of radio emission from galactic and extragalactic sources, and the wavelength dependence of radio source size.

If the absorption cross-section is constant then the energy distribution of atoms has a form

$$N_2/N_1 = const(E_2 - E_1)kT. \qquad (11)$$

For all real values of the temperature and wavelength of radiation the ratio (11) is much smaller than unit. For example, at $\lambda = 1m$ the ratio (11) is less than 1, if $T < 10^{18}$K. It implies that $N_1 \approx N$, where $N$ is the total number of atoms. This relation was used in [5] to obtain the condition for emission. There are no reasons to expect that the Boltzmann distribution is still valid for an ensemble of atoms interacting with thermal radiation.

From the relation $B_{21}B_\kappa \approx 1$ we obtain the stimulated emission cross-section in the form

$$\sigma_{21} \approx \hbar\omega / (\pi B_\nu) = \lambda l_T / (2\pi), \tag{12}$$

where $\lambda$ is the wavelength, and $l_T = 2\pi\hbar c/(kT)$. Thus, Einstein's relation $B_{12} = B_{21}$, equivalent to $\sigma_{12} = \sigma_{21}$, is not valid in the Rayleigh-Jeans region.

Consider now the Wien region $\hbar\omega > kT$. There are strong theoretical and observational indications of the stimulated character of thermal blackbody radiation in the whole range of spectrum [8]. First, the spectral energy density of thermal blackbody radiation is described by a single Planck's function at all frequencies. Second, there are clear observational indications of the existence of thermal harmonics in stellar spectra and of laser type sources. The latter are connected with the induced origin of thermal radiation, similarly to well known maser sources. In particular, a possible non-saturated X-ray laser source emitting in the Fe $K_\alpha$ line at 6.49 keV was recently discovered in the radio-loud quasar MG J0414+0534 [9].

Assuming the stimulated origin of thermal blackbody radiation, i.e. $\nu_s < \nu_i$, from Eq.(8) one can obtain the relation $B_{21}B_\nu = \sigma_{21}\pi B_\nu / (\hbar\omega) \approx 1$, and then the stimulated emission cross-section in the form

$$\sigma_{21} \approx (\lambda^2 / 2\pi) \exp(\hbar\omega / kT). \tag{13}$$

The absorption cross-section is likely to be constant in the whole range of spectrum, so the ratio $N_2 / N_1$, according to Eq.(6), is given by the formula

$$N_2 / N_1 \approx (\sigma_{12}\omega^2 / 2\pi c^2) \exp(-\hbar\omega / kT), \tag{14}$$

which is somewhat similar to the Boltzmann law.

The exact formula for the ratio $N_2 / N_1$ can be obtained from Eq.(6):

$$N_2 / N_1 = \sigma_{12}\omega^2 / (2\pi c^2)(\exp(\hbar\omega / kT) - 1). \tag{15}$$

The function (15) has a maximum at $\hbar\omega = 1.6kT$. It means that, in the field of thermal blackbody radiation, the excited levels with $E_2 - E_1 \approx kT$ are the most populated. Numerically, $(N_2 / N_1)_{max} = 0.8 \times 10^{-15} T^2$, so the ratio $N_2 / N_1$ is less than 1, if only $T<3*10^7$ K. For the temperatures $T>4*10^7$ K the population of the levels corresponding to the maximum of the function (15) is inverse. It suggests that laser type sources are more easily realized in X-ray and gamma-ray range of spectrum than in the optical region. The well known example is a possible gamma-laser in the Galactic Center emitting in the line 0.511 MeV [10].

The author is grateful to A.V.Postnikov for useful discussions.